\documentclass[12pt,preprint]{aastex}
\usepackage{times,color,ulem}
\newcommand{\sdo}{{\it SDO}}

\shorttitle{White-Light Limb Flare}
\shortauthors{Zhao et al.}

\begin{document}
\title{White-light Continuum Observation of the Off-limb Loops of the 
SOL2017-09-10 X8.2 Flare: Temporal and Spatial Variations}

\author{Junwei Zhao\altaffilmark{1}, Wei Liu\altaffilmark{2,3}, and 
Jean-Claude Vial\altaffilmark{4}}
\altaffiltext{1}{W.~W.~Hansen Experimental Physics Laboratory, Stanford
University, Stanford, CA 94305-4085, USA}
\altaffiltext{2}{Lockheed Martin Solar and Astrophysics Laboratory, Palo Alto,
CA94304, USA}
\altaffiltext{3}{Bay Area Environmental Research Institute, NASA Research Park, 
Moffett Field, CA 94035, USA}
\altaffiltext{4}{Institut d'Astrophysique Spatiale, Universit\'{e} 
Paris-Saclay, C.N.R.S., Batiment 121, ORSAY 91405, France}

\begin{abstract}
Observations of the Sun's off-limb white-light (WL) flares offer rare
opportunities to study the energy release and transport mechanisms in flare 
loops. One of the best such events was SOL2017-09-10, an X8.2 flare that
occurred near the Sun's west limb on 2017 September 10 and produced a WL
loop system lasting more than 60 minutes and reaching an altitude higher 
than 30 Mm. The event was well observed by a suite of ground- and 
space-based instruments, including the {\it Solar Dynamics 
Observatory}/Helioseismic and Magnetic Imager (\sdo/HMI) that captured
its off-limb loops in WL continuum near \ion{Fe}{1} 6173 \AA, 
and the Atmospheric Imager Assembly (\sdo/AIA) that observed its 
ultraviolet (UV) and extreme-ultraviolet (EUV) counterparts. We found
quasi-periodic pulsations in the WL and UV emissions at the flare 
loop-top with a period around 8.0 min. Each pulsation appears to have 
an EUV counterpart that occurs earlier in time and higher in altitude. 
Despite many similarities in the WL and 
UV images and light curves, the WL flux at the loop-top continues 
to grow for about 16 minutes while the UV fluxes gradually decay.
We discuss the implication of these unprecedented observations on the 
understanding of the enigmatic off-limb WL flare emission mechanisms.
\end{abstract}


%\newpage

\section{Introduction}
\label{sec1} 

White-light flares \citep[WLFs;][]{hud06, hud11} are rare flaring events that cause 
continuum emission in excess of the photospheric background. Off-limb WLFs, 
with flare-loop brightening observed in white light (WL) beyond the solar 
limb, are even rarer with only a few cases reported \citep[e.g.,][]{hie92, 
mar14}. Three off-limb WLFs were so far observed in the continuum 
near \ion{Fe}{1}~6173~\AA\ by the Helioseismic and Magnetic Imager 
\citep[HMI;][]{sch12a, sch12b} onboard the spacecraft {\it Solar Dynamics 
Observatory} \citep[\sdo;][]{pes12}, i.e., SOL2013-05-13T02:16 (X1.7), 
SOL2013-05-13T16:01 (X2.8), and SOL2017-09-10T15:35 (X8.2). These observations 
offered unprecedented opportunities for solar physicists to study properties 
of WLFs that are otherwise impossible: e.g., \citet{mar14} compared 
the WL brightening sources with those observed in extreme ultraviolet (EUV) 
and soft X-ray; \citet{sai14} analyzed the linear polarization observed in 
the WL flare loops; and \citet{hei17} studied emission mechanisms of the 
WL off the solar limb.

One popular phenomenon that is associated with solar flaring events and has 
been widely studied by various authors is quasi-periodic pulsations (QPPs) 
in the time-dependent intensity curves observed in different wavelengths 
during flares \citep{nak09, van16}. This phenomenon, with a repeating 
period of a few seconds to a few minutes, has been observed almost across 
all the electromagnetic spectrum, such as in $\gamma$-rays \citep{nak10}, 
EUV and soft X-rays \citep[e.g.,][]{den17, dom18}, chromospheric lines 
\citep[e.g.,][]{bro16, tia16}, and also in radio bursts \citep[e.g.,][]{li15}. 
Due to the rare observation of WLFs, no known QPP has yet been reported 
in white light. The physical mechanism for the generation of QPP is not 
very clear, but it is believed that this phenomenon is either due to the 
characteristics of the flaring loops that exhibit magnetohydrodynamic 
(MHD) oscillations at the flaring sites or MHD waves passing through the 
flaring loops \citep[e.g.,][]{nak09}, or due to the quasi-periodic magnetic 
reconnection and energy release that power the intensity enhancements 
of the spectrum lines \citep[e.g.,][]{mur09}. 

The X8.2 limb flare of 2017 September 10 (SOL2017-09-10T15:35), whose loops
rose right above the Sun's west limb and which was well observed by a suite 
of ground- and space-based instruments, has been studied from many 
different perspectives to tackle various physics problems related to 
flare dynamics and energy releases. For example, using EUV observations 
\citet{war18} studied the formation and evolution of the current sheet 
following the eruption; Coupling hard X-ray and ground-based radio observations, 
\citet{gar18} studied the evolving spatial and energy distribution of 
high-energy electrons in the flaring region; Coupling with {\it GOES} X-ray 
and EUV observations, \citet{hay19} found QPPs with 
periods of about 65 s and 150 s; Using the {\sdo}/HMI's WL observations, 
\citet{jej18} modeled the physical conditions inside the flare loops 
and explored the WL emission mechanisms; \citet{hei20} identified 
signatures of He continuum in the cool flare loops; Using \ion{Ca}{2} 
Stokes components observed by the Swedish 1m Solar Telescope, \citet{kur19} 
reported a strong magnetic field of 350~G at the flare's loop-top; Using 
microwave observations, \citet{fle20} estimated that the coronal magnetic 
field in the flaring loops decayed with a rate of $\sim5$~G~s$^{-1}$ during 
the flare, and \citet{yu20} analyzed the bidirectional outflows from the 
magnetic reconnection sites. 

In this Letter, combining simultaneous observations of the SOL2017-09-10 
flare obtained in the \sdo/HMI's continuum intensity and \sdo/AIA's 
\citep[Atmospheric 
Imaging Assembly;][]{lem12} ultraviolet (UV) and EUV observations, along 
with data from the {\it Geostationary Operational Environmental Satellite} 
({\it GOES}) and {\it Reuven Ramaty High Energy Solar Spectroscopic 
Imager} \citep[{\it RHESSI};][]{lin02}, we analyze the temporal and spatial 
variations of the brightening observed in different wavelength channels, 
as well as the QPPs in all these channels. It is believed that these 
unprecedented simultaneous observations in different wavelength channels
offer us a rare opportunity to understand the
emission mechanism of white light, heat dissipation process, and energy 
releases during the flare. This Letter is organized as follows: we introduce 
observations in Section 2, present our analysis and results in Section 3, 
and discuss our results in Section 4.

\section{Observations}
\label{sec2}

\begin{figure}[!t]
\epsscale{1.00}
\plotone{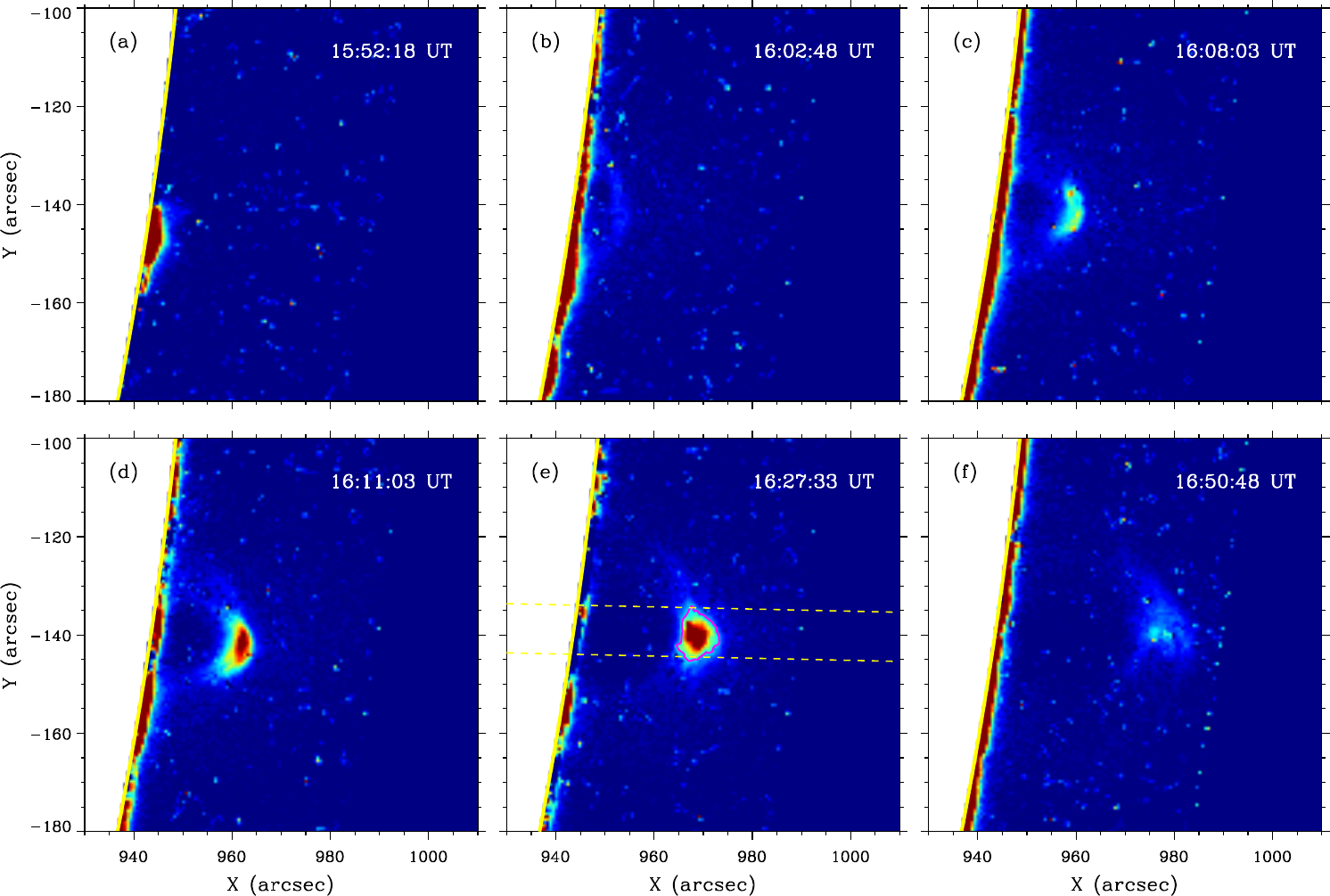}
\caption{Selected white-light images from the \sdo/HMI observations show
temporal evolution of the flare's off-limb loops, after a 
long-time-averaged pre-flare background is removed. White area on the left 
side of each panel indicates the Sun's photospheric disk and limb. Yellow 
dashed lines in panel (e) delimit the area inside which intensities are 
integrated for further analysis, and the magenta contour showing the WL
loop-top is used in Figure~\ref{AIA_img}c for a location comparison. (An 
animation showing the entire sequence of the event is available online,
with a total running time of 7 seconds.) }
\label{HMI_img}
\end{figure}

The \sdo/HMI uses \ion{Fe}{1} 6173~\AA\ line to observe the Sun with 1\arcsec\ 
spatial resolution and 45 sec temporal cadence, and provides full-disk 
continuum intensity data, along with Doppler velocity and magnetic field.
The field of view of \sdo/HMI is nominally $45\arcsec$ wider than the 
photospheric 
solar disk, allowing detection of features bright in visible light immediately 
beyond the solar limb. The SOL2017-09-10 X8.2 flare started at approximately 
15:35~UT in NOAA Active Region 12673, when most of the active region rotated 
around the west limb into far side of the Sun. The sudden brightening at the 
footpoints of the flare in the chromosphere and the rise of the bright
flare loops, up to about $40\arcsec$ above the limb, were observed by the 
\sdo/HMI in its continuum intensity, i.e., white light. As shown in 
Figure~\ref{HMI_img} and its associated online animation, the flare loops 
become visible above the limb at around 16:03:00~UT. Its loop-top brightness 
gradually enhances with the increase of height, reaching an intensity maximum 
at about 16:12:00~UT when the \sdo/AIA's UV channels also reach their 
respective maxima (Cf. Figure~\ref{light_curve}b). While continuing to rise 
in height, the WL intensity drops for about 5 min, and then starts to 
enhance again, reaching a even brighter maximum at about 16:27:45~UT. The 
WL brightening of the flare loops then gradually fade away, lasting a total 
of about 70 min.

\begin{figure}[!t]
\epsscale{1.00}
\plotone{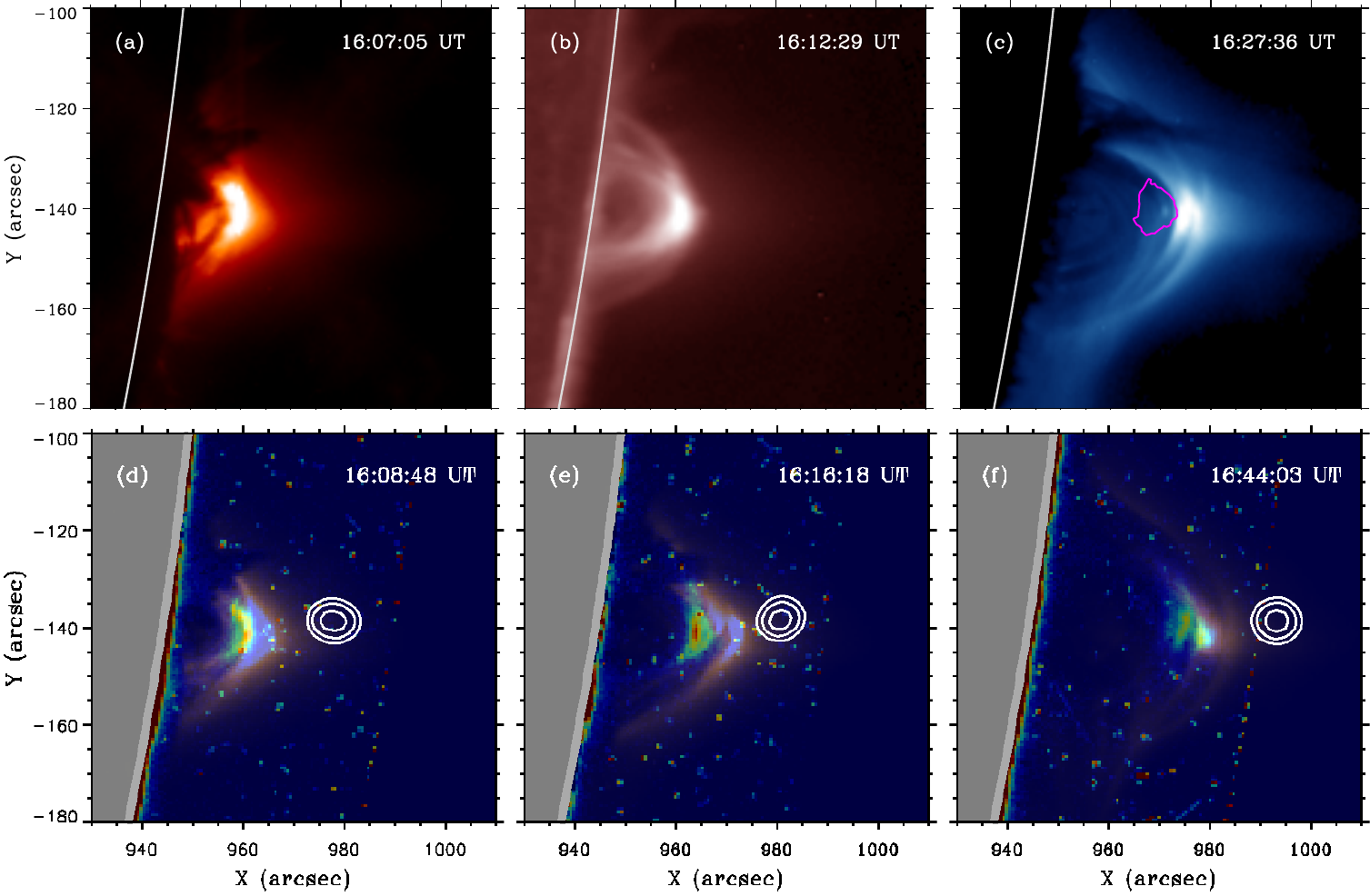}
\caption{Selected images of the flare loops at the wavelength channels of 
(a) 304~\AA\ near its peak intensity at 16:07:05~UT; (b) 1700~\AA\ near
its peak intensity at 16:12:29UT; and (c) 335~\AA\ at 16:27:36UT, 
overplotted with a magenta WL contour near the WL peak time (see 
Figure~\ref{HMI_img}e).  (d)-(e) Selected composite images 
of HMI WL (blue-red background image) and AIA 335~\AA\ (orange-white 
semi-transparent foreground image), with {\it RHESSI} 10-16~keV intensity 
over-plotted as white contours.}
\label{AIA_img}
\end{figure}

The flare was also observed in all the \sdo/AIA wavelength channels. However, 
due to the powerful intensity of this flare, a number of AIA channels either
saturated near and during the peak of the flare, mostly at the location of 
the loop-top, or had irregular exposure times in order not to saturate the 
observations. The data from those channels with irregular exposure times may 
still be useful for some types of analyses, but cannot be used in this 
analysis, 
which focuses on temporal variations of the loop-top intensity and is sensitive 
to the alteration of exposure times. This leaves only observations from the 
wavelengths of two UV channels (1700~\AA\ and 1600~\AA) and two EUV channels 
(335~\AA\ and 304~\AA) useful for the analysis in this study. According to 
\citet{lem12}, the UV channel 1700~\AA\ (1600~\AA) is sensitive to the 
temperature of $10^{3.7}$~K ($10^{5.0}$~K) corresponding to the upper 
photosphere (the upper photosphere and transition region), the EUV channel 
304~\AA\ (335~\AA) is sensitive to the temperature of $10^{4.7}$~K 
($10^{6.4}$~K) corresponding to the chromosphere and transition region (active
region corona). However, it must be stressed that these properties of 
different channels are assessed mostly for on-disk observations, and may
not strictly apply to the off-limb observations. 
Figure~\ref{AIA_img} shows key snapshots of the flare loops for 
some wavelengths, along with selected composite images showing the relative
spatial and temporal relations of WL and EUV channels. Particularly, as 
seen in Figure~\ref{AIA_img}c, the WL loop-top, at its peak brightness, 
is located below the 335\AA\ loop-top in its absorption area. 

{\it GOES} captured the total soft X-ray flux of the Sun in 
its both channels: 1--8~\AA\ and 0.5--4~\AA\ (Figure~\ref{light_curve}a). 
{\it RHESSI}'s X-ray images did not cover the entire flare event, but 
its coverage during the early impulsive phase of the flare provides
useful information that can be compared with the WL/UV/EUV locations 
(Figure~\ref{time_ht}f). Selected {\it RHESSI} X-ray images are also 
plotted as contours with the WL and EUV images for comparison of 
loop-top locations (Figure~\ref{AIA_img}d-f).

\section{Results}
\label{sec3}

\subsection{Time--Height Relation}
\label{sec31}

\begin{figure}[!t]
\epsscale{1.00}
\plotone{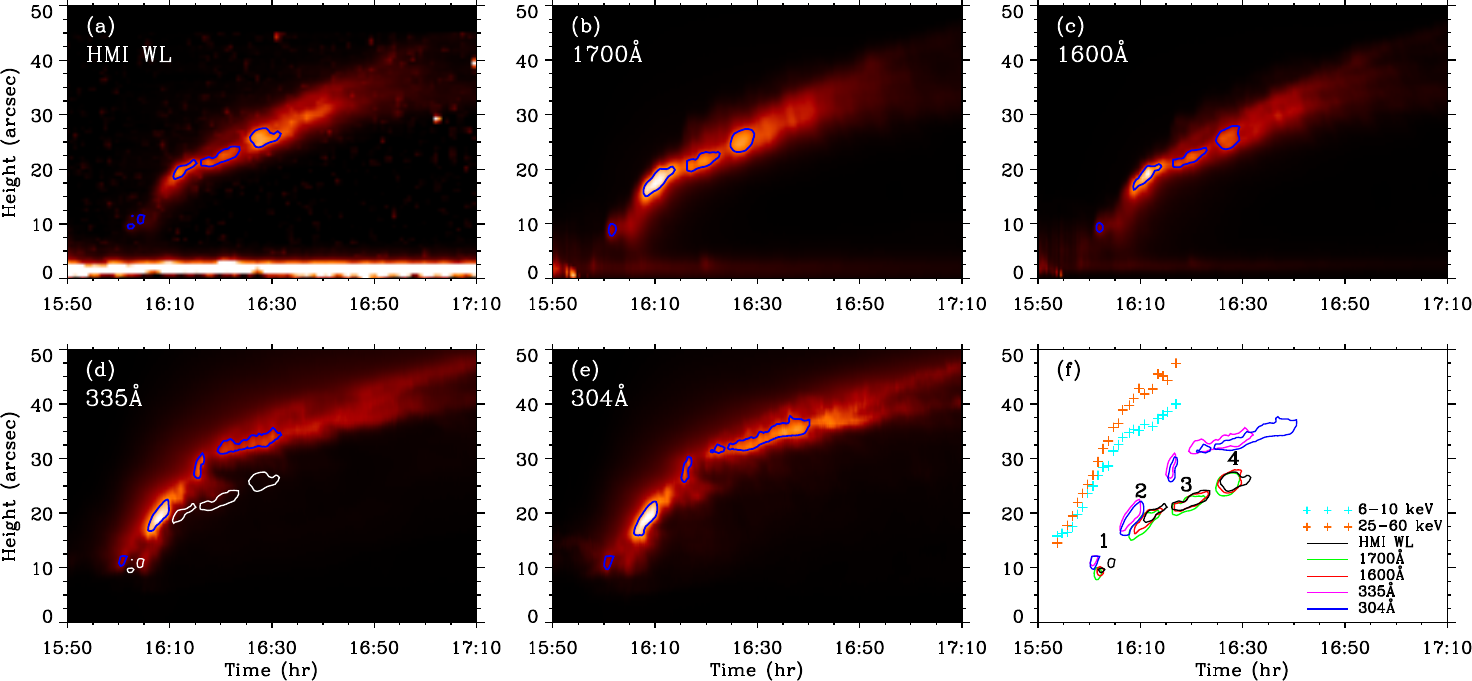}
\caption{Time--height relations obtained for the flare's loop-top after an 
integration over a $10\arcsec$-wide band for each time step, using
observations from (a) HMI WL, (b) 1700~\AA, (c) 1600~\AA, (d) 335~\AA, and 
(e) 304~\AA. Contours in each of these images mark the 75\% of intensity 
level at the peak of their corresponding brightness enhancement patches.
The WL contours are also plotted in panel (d) to better 
compare the WL emission relative to EUV emissions. (f) Contours in panels 
(a) -- (e) are plotted together and numbers ``1" -- ``4" are marked 
corresponding to the four brightening patches. The centroid locations 
obtained from {\it RHESSI}'s 6-10~keV and 25-60~keV soft X-ray data are 
also overplotted as `+' signs for a comparison. }
\label{time_ht}
\end{figure}

To analyze the time--height relations of the loop-top observed in all the five
visible, UV, and EUV channels through the period of interest from 15:56~UT 
to 17:12~UT, for each time step and for each altitude we integrate the 
intensity for the total flux in a $10\arcsec$-wide band between 
the dashed lines marked in 
Figure~\ref{HMI_img}d, which best includes the brightest part of the 
loop-top. The integrated brightness for all the five wavelength channels,
as well as their comparison, are shown in Figure~\ref{time_ht} as a function 
of time and height above the photospheric limb. Note that 
the HMI, AIA's UV and EUV channels have different temporal cadences of 
45~s, 24~s, and 12~s, and HMI and AIA have different pixel sizes of 
about $0.5\arcsec$ and $0.6\arcsec$; therefore, the time--height plots 
in Figure~\ref{time_ht} are shown after all the AIA images are interpolated 
to match the WL image resolution and are aligned with the WL images in 
space and time.

It can be seen from Figure~\ref{time_ht} that through all the five wavelength
channels, the loop-top shows four distinct brightness-enhancements patches, 
each of which is marked by one contour. Within each wavelength channel, 
every brightness-enhancement patch differs each other in both enhancement 
amount and duration, therefore we plot the contours using 75\% of the 
peak intensities of their respective patches. As shown in Figure~\ref{time_ht}f,
based on the time and location of the brightness-enhancement patches, the 
five wavelength channels form two groups, 
with one group being the WL and UV channels (1700~\AA\ and 1600~\AA), and 
another group the EUV channels (335~\AA\ and 304~\AA). Within each group, 
the height and the duration of the four brightness enhancements observed in 
different channels are quite similar but do not exactly overlap. The brightness 
enhancements in the EUV channels occur earlier in time and higher in altitude 
than those in the WL and UV channels, and the time lags and height differences 
between the two groups of enhancement patches increase as the flare loop-top 
evolves to higher altitude. The X-ray centroid locations of the loop-top, 
obtained from the {\it RHESSI}'s 6-10~keV and 25-60~keV channels during 15:53UT 
and 16:17UT, can be found at least 10\arcsec\ higher and at least 10 minutes 
earlier than the AIA's EUV channels (Figure~\ref{time_ht}f).

\begin{figure}[!t]
\epsscale{0.5}
\plotone{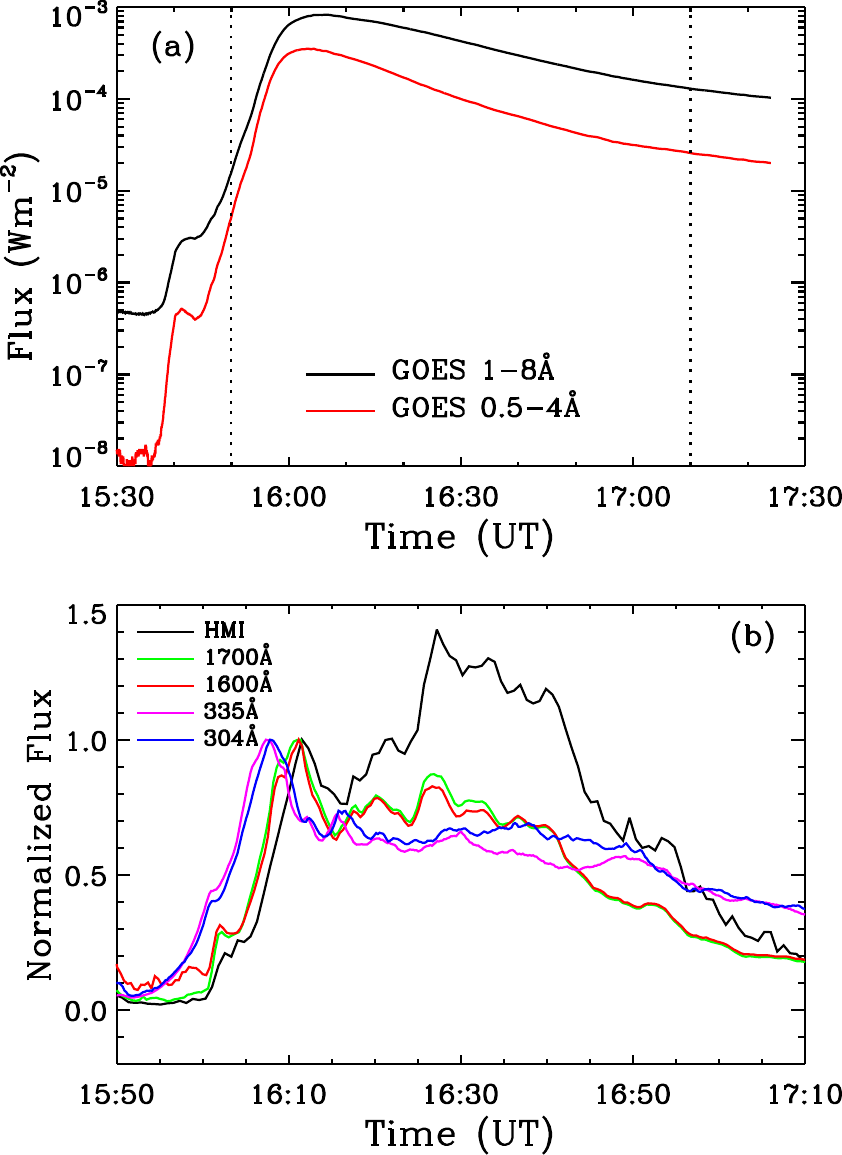}
\caption{(a) Soft X-ray flux from {\it GOES} during the flare's impulsive and 
gradual phases. Vertical dotted lines indicate the time window
for the curves plotted in panel (b). (b) Temporal evolution of the 
flux for the HMI and AIA wavelength channels, integrated in the stripe shown
in Figure~\ref{HMI_img}e. All fluxes are normalized relative to their 
respective peak fluxes during 16:05 -- 16:15UT.} 
\label{light_curve}
\end{figure}

To better compare the temporal variations of the intensity for all the 
WL/UV/EUV channels, we further integrate the time--height diagrams over 
heights, and study the total 
flux change over time. In order to focus only on the temporal changes in 
the loop-top, we integrate the intensity between the heights of $6\arcsec$ and 
$45\arcsec$ above the limb, excluding the upper photosphere that keeps elevated
brightness throughout the analyzed period in the WL and UV channels.
Figure~\ref{light_curve}b shows all the flux curves, which are 
normalized to 1.0 at their respective major brightness peak between 16:05 
-- 16:15~UT.  It can  be seen that all the UV and EUV channels show similar 
trend of growth and decay, despite that the UV channels are grouped together 
with the WL in the brightness-enhancement time and altitude. Showing 
a growth trend similar to the UV channels before the first major peak 
around 16:12~UT, the WL light flux continues to grow to a brighter peak 
around 16:28~UT after a 5-min dip in brightness following the first peak. 
This can also well be seen in Figure~\ref{HMI_img} and its accompanied animation. 
That is, the WL and UV channels show sharply different behaviors after about 
16:17~UT despite their many similarities in other aspects, highlighting the 
different emission mechanisms in the WL and UV wavelengths at the flare's 
loop-top. For comparison, the {\it GOES} soft X-ray flux is plotted in 
Figure~\ref{light_curve}a. We need to keep in mind that the {\it GOES} flux 
is an integration of the X-ray flux of the entire Sun, including the flare's 
footpoints, loops, and elsewhere, while the fluxes shown in 
Figure~\ref{light_curve}b are obtained only from the flare's off-limb loop-top.

\subsection{Quasi-Periodic Pulsations}
\label{sec32}

\begin{figure}[!b]
\epsscale{0.85}
\plotone{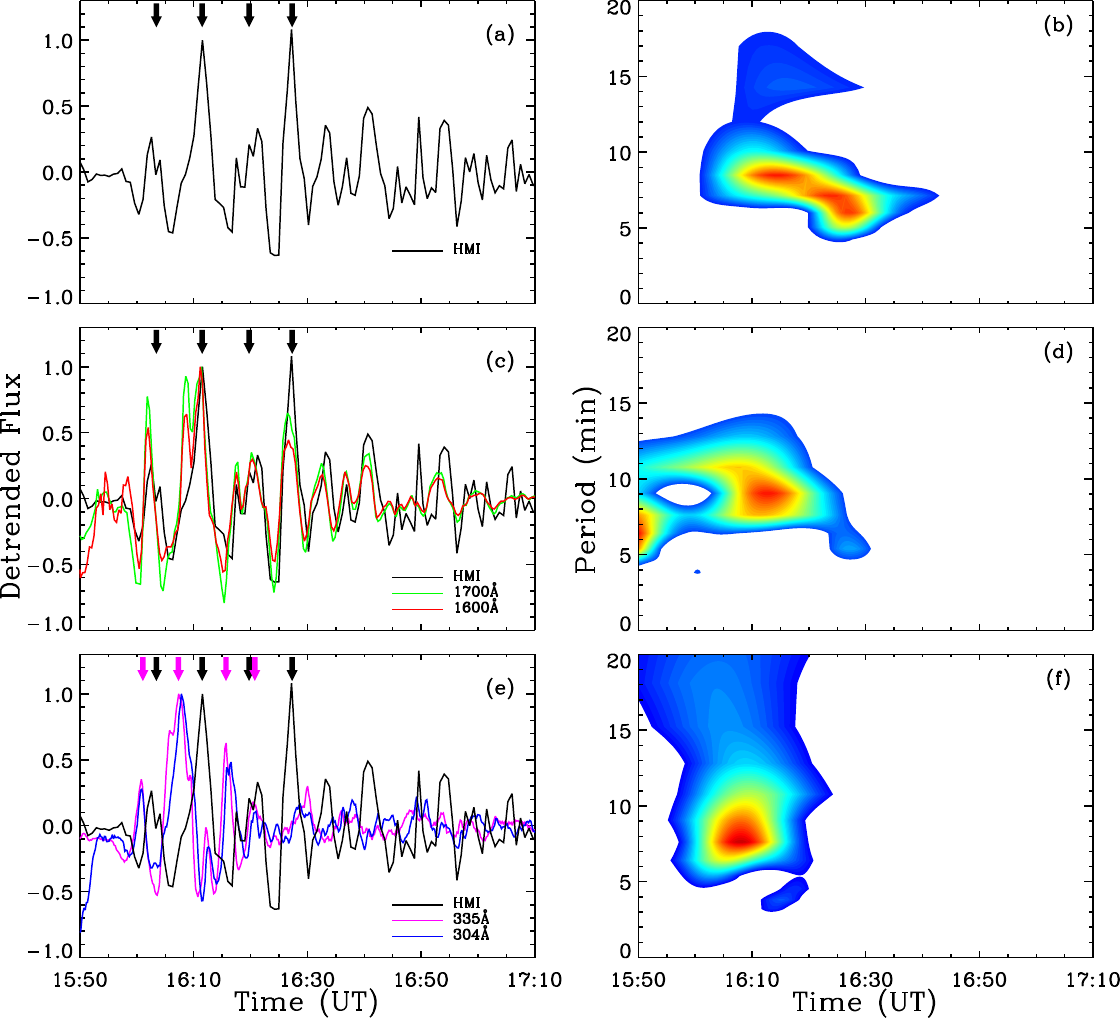}
\caption{(a) Detrended HMI WL flux curve, and (b) wavelet power of the 
detrended WL curve. (c) Comparison of detrended flux curves between 
the WL and the UV channels, and (d) wavelet power of the detrended 1600~\AA\ 
curve. (e) Comparison of detrended flux curves between the WL and the 
EUV channels, and (f) wavelet power of the detrended 335~\AA\ curve. Dark 
arrows in the left panels point to the four peak times in the detrended 
WL curve, and the magenta arrows indicate the peak times of the detrended 
335~\AA\ curve. The shown wavelet power maps are part of a map that 
has a longer time sequence of 2 hr with a longer period range, thus the 
influence cones locate outside of the plotted range. } 
\label{curves}
\end{figure}

To study the shorter-period pulsations in the brightness observed in all the 
wavelengths, we detrend the brightness curves shown in 
Figure~\ref{light_curve}b by subtracting their respective running averages 
with a window width of 300 sec, a practice commonly used in studying 
quasi-periodic pulsations \citep[e.g.,][]{hay19, yua19}. We have experimented
with different window widths, and found that the 300-sec window width 
only suppresses long-period pulsations with little impact on the 
short-period pulsations that are of interest in this study. The left panels of 
Figure~\ref{curves} show the detrended flux curve of the WL
and its comparisons with the detrended curves from the UV and EUV channels,
and the right panels of Figure~\ref{curves} show the wavelet power of the 
detrended curves of the WL, one UV channel (1600~\AA), and one EUV channel 
(335~\AA).

It can be seen that both the WL and UV channels show four near-simultaneous
brightness peaks, with a seemingly one-to-one correspondence to the four 
brightness-enhancement patches shown in Figure~\ref{time_ht}. The WL brightness 
peaks, marked by the dark arrows, show quasi-periodic pulsations with a period
of approximately 8.0~min. The pulsations in the two UV channels occurred 
slightly earlier than those in the WL. The wavelet power calculated 
from the detrended WL flux curves peaks around the period of 8.0 min at 
the beginning, and gradually shifts to the period of 5 min, lasting a total of 
about 40 min. The wavelet power calculated from the 1600~\AA\ shows a wider
period range between $\sim$7 and $\sim$11 min and lasts for about 30 min.

The pulsations in the two EUV channels, indicated by the four magenta arrows in 
Figure~\ref{curves}e, lead the WL pulsations by $\sim$2.4 -- 6.5~min. 
%{\color{red} \sout{It seems reasonable 
%to believe that the four EUV pulsations also correspond to the four EUV 
%brightness-enhancement patches, and the four EUV pulsations have a one-to-one 
%correspondence to the four WL pulsations.}} 
%Compared with the WL pulsations, the 
These EUV pulsations are less regular in periodicity with an averaged period 
of $\sim$6.6~min, significantly shorter than that of the WL pulsations. 
Between the two EUV channels, the pulsations in 335~\AA\ channel lead those 
in 304~\AA\ channel slightly by about 28 sec, estimated through 
cross-correlation. 
It is rather interesting that there are four episodes of emission enhancements across EUV, UV, and WL channels. However, it is not clear whether there exists an one-to-one physical correspondence among the pulsations of these channels, because of their different emission mechanisms and the complex flare structure involving an arcade of loops distributed along the line-of-sight that can contribute to the EUV emission due to its optically-thin nature.
%{\color{red} Because of the complicated structure of the flare loop 
%and the optically-thin nature of UV/EUV observations, as well as different 
%emission mechansisms between WL and UV/WUV, it is not straightforward whether 
%these EUV pulsations have clear one-to-one correspondence with the WL/UV 
%pulsations.}

\section{Discussion}
\label{sec4}

We have analyzed the rare white-light observation of the evolution 
of an off-limb loop system in the SOL2017-09-10 X8.2 flare, and found that
(1) there are quasi-periodic pulsations in the WL and UV channels with a period 
of around 8.0~min, and less regular pulsations with an average period of 
6.6~min in the EUV channels; (2) intensity-enhancements appear sequentially
in WL and UV/EUV channels with time delays and altitude decreases corresponding
to progressively lower temperatures of the emitting plasma; and (3) the WL 
flux of the flare's loop-top continues to grow for about 16 more 
minutes while the UV/EUV fluxes decay.

The quasi-periodic pulsations observed across the WL/UV/EUV channels, with
different periods and irregular time delays between different channels, implies
that these pulsations are unlikely of the MHD wave nature or oscillatory
properties of the flaring plasma. Through analyzing the evolution of 
a strong on-disk flare, \citet{asa04} suggested that repeated enhancements 
of X-ray/EUV emissions were indicative of energy release and particle 
acceleration episodes resulting from magnetic reconnection. Our observation 
of the limb-flare loop presents a different view angle of a similar 
phenomenon, whose quasi-periodic brightenings are also possibly caused by 
episodic magnetic reconnection above. For this same flaring event, \citet{ree21}
found damped oscillations in Doppler velocity with a period $\sim$400 s in
the region south of the the main flare arcade, and 
suggested that the oscillations are caused by magnetic reconnection
outflows. Despite different areas of the same event analyzed, the loop-top of 
the main arcade in our study and the loop-top south of the main arcade 
in theirs, the similarities in the pulsation periods may point to a 
common physical cause -- pulsed magnetic reconnection.

The brightness-enhancement patches across different wavelengths can 
serve as tracers of the flare's energy release and subsequent thermodynamic 
evolution across different temperature and density ranges. As 
Figure~\ref{time_ht}f shows, the emission generally progresses with time 
toward lower altitudes from X-ray to EUV and then UV/WL, 
corresponding to the emitting plasma at progressively lower temperatures.
This is a well-known trend that can result from the interplay of (a) 
upward development of magnetic reconnection toward higher altitude with
time in the standard flare model, (b) downward contraction of newly
reconnected hot flaring loops that undergo cooling at the same time.
Such contractions have various observational 
manifestations, such as shrinkage of flare loops \citep{sve87} at speeds 
of the order of 10~km~s$^{-1}$ seen in soft X-rays \citep{for96, ree08}, and
the so-called supra-arcade downflows \citep[e.g.,][]{mck99, sav11, liu13}. 
In our analysis (see Figure~\ref{time_ht}f), the altitudes for the first 
two groups of brightenings, around 16:02UT and 16:10UT across all 
wavelengths, are similar at about 10\arcsec\ and 18\arcsec\ above 
the limb, while the time delays between the EUV and UV emissions are 
about 2 and 4~min, respectively.
For the other two groups of later brightenings, the EUV emission originates
about 7\arcsec\ higher than the UV/WL emission with time delays of 3 -- 
5 min, which translates to a speed of $\sim 17 - 28$ km~s$^{-1}$ if
we assume this results from loop shrinkage. This trend agrees with the 
increase with time of the separation between 
the 6-10 and 25-60 keV X-ray emission centroids \citep[especially after
about 16:05UT; cf.,][]{liu04}. 

Despite many similarities in WL and UV observations, their integrated 
light curves show different trends: the WL flux continues to grow 
while the UV flux starts to decay after the peak around 16:10UT. This 
implies, unsurprisingly, that the UV and WL have different emission mechanisms. 
Off-limb flare loops are visible in WL primarily because of a combination 
of the following processes: hydrogen Paschen and Brackett recombination, 
hydrogen free-free continuum emission, and Thomson scattering of the solar 
radiation \citep{hei17, jej18}. It is likely that before 16:10UT plasma 
temperature is relatively high at the flare's loop-top, and all the WL emission 
mechanisms play a role. After 16:10UT and above the height of approximately 
$22\arcsec$ (see Figure~\ref{time_ht}a), the temperature may have dropped
substantially and resulted in reduced UV/EUV emissions; however, the increasing 
electron density may still keep the WL intensity growing for a longer time, 
primarily because the Paschen and Brackett recombination and free-free 
emission depend quadratically on the density. However, one should also 
realize that high densities also correspond to non-negligible opacity
leading to a decrease of emission. In fact, a detailed comparison of WL 
and EUV~335\AA\ emissions in Figures~\ref{AIA_img}c and \ref{time_ht}d, 
along with the sequence of 335\AA\ images, shows that during the EUV 
decay period, the WL emissions are located in an area of reduced 335~\AA\ 
emission that appears to be a dark absorption. This is also evident on other 
AIA EUV channels, such as 193\AA\ and 211\AA\ (cf. Figure 1 of Heinzel 
et al. 2020 and Figure 3 of Jej\v{c}i\v{c} et al. 2018), providing 
another piece of evidence that plasma has cooled substantially 
in the areas where WL intensity enhances. 

%Overall, our analysis of the WL observations of the off-limb loops reveals 
%quasi-periodic pulsations at the loop-top with a period around 8.0~min, which
%are similar to those observed in the UV channels, but lag behind and occur 
%lower in altitude than those in the EUV channels. This implies that the 
%pulsations likely originate from a higher source rather than being a display 
%of plasma's oscillatory property. That the WL and UV show different intensity 
%evolution indicates that different WL emission mechanisms play leading roles 
%below and above the height of 22\arcsec.

\acknowledgments We thank an anonymous referee for constructive comments that
helped shape this paper to the current version. Both {\it SDO} and {\it 
RHESSI} are NASA space missions, and {\it GOES} are NOAA satellites.
J. Z. is partly sponsored by NASA grant 80NSSC18K0668. W. L. was supported by NASA grants NNX16AF78G, 80NSSC21K1327, and 80NSSC21K1687.

\end{document}